
%
%

\input harvmac.tex

\def\frac#1#2{{\textstyle{#1\over #2}}}
\def\Tn#1{({\buildrel \leftarrow\over{T^{#1}}})}

\Title{PUPT--1300/ LAVAL-PHY-28/91}{\vbox {
\centerline{Integrability of the Quantum KdV equation}
\bigskip
\centerline{at $c = -2$.}
}}

\centerline{P. Di Francesco\foot{Work supported by NSF grant
PHY-8512793.}}
\bigskip\centerline
{\it Joseph Henry Laboratories,}
\centerline{\it Princeton University,}
\centerline{\it Princeton, NJ 08544.}
\bigskip
\centerline{and}
\bigskip
\centerline{P. Mathieu\foot{Work supported by NSERC (Canada) and
FCAR (Qu\'ebec).} and D. S\'en\'echal\foot{Work supported by
NSERC (Canada)}}
\bigskip\centerline{\it D\'epartement de Physique,}
\centerline{\it Universit\'e Laval,}
\centerline{\it Qu\'ebec, Canada, G1K 7P4.}
\vskip .2in

\noindent
We present a simple a direct proof of the complete integrability of the
quantum KdV equation at $c=-2$, with an explicit description of all the
conservation laws.

\Date{12/91}


The seminal work of Zamolodchikov
\ref\ZAMO{A.B. Zamolodchikov, JETP Lett. {\bf 46} (1987) 160;
Int. Jour. Mod. Phys. {\bf A3} (1988) 4235; Adv. Stud. in Pure Math.(1989).}
on integrable perturbed minimal models has brought to light the relevance of
quantum KdV (qKdV) equations in conformal field theory
\ref\SAYA{R. Sasaki and I. Yamanaka, Adv. Stud. in Pure Math. {\bf 16}
(1988) 271.}\ref\KM{B.A. Kuperschmidt and P. Mathieu, Phys. Lett. {\bf B227}
(1989) 245.}.
For instance the integrals of motion which are preserved by the $\phi_{(1,3)}$
perturbation are known to be exactly equivalent to the qKdV conservation laws
\ZAMO\KM\ref\EY{T. Eguchi and S.K. Yang, Phys. Lett. {\bf B224} (1989) 373;
T.J. Hollowood and P. Mansfield, Phys. Lett. {\bf B226} (1989) 73.}.
(By `qKdV equation' we mean the KdV equation formulated canonically via the
quantum form of its second Hamiltonian structure).
The full integrability of the qKdV equation has been widely expected but
proven rigorously only very recently by Feigin and Frenkel
\ref\FF{B. Feigin and E. Frenkel, {\it Free field resolution in affine Toda
field theories}, RIMS-Harvard preprint (october 1991).}.
By using the relationship between the qKdV equation and the quantum
sine-Gordon equation they have been able to establish the existence of an
infinite number of mutually commuting conservation laws.
However, their argument is not constructive, and it remains an interesting
problem to find an explicit form for these conservation laws.
We already know that one cannot expect a recursive description of the
commutation laws \`a la L\'enard since the qKdV equation is not
bi-Hamiltonian\KM.

A step in this direction was made in
\ref\DiMA{P. Di Francesco and P. Mathieu, PUPT-1284/LAVAL-PHY-26/91 (september
1991).}\ref\EYKNS{T. Eguchi and S.K. Yang, Phys. Lett. {\bf B335} (1990) 282;
A. Kuniba, T. Nakanishi and J. Suzuki, Nucl. Phys. {\bf B356} (1991) 750;}
where it was shown that the conservation law of dimension $2k-1$ is exactly
proportional to the vacuum singular vector of the non-unitary minimal models
with $(p,p')=(2,2k+1)$, \foot{This was first observed in
\ref\FREUND{P.G.O Freund, T.R. Klassen and E. Melzer, Phys. Lett.
{\bf B229} (1989) 243.}.} whose expression is
known explicitly \ref\BSA{L. Benoit and Y. Saint-Aubin, Phys. Lett. {\bf B215}
(1988) 517; M. Bauer, P. Di Francesco, C. Itzykson and J.B. Zuber,
Nucl. Phys. {\bf B362} (1991) 515.}.

Another result is presented here: we display the explicit expression of
all conservation laws at $c=-2$ and prove their commutativity. The form of
these conservation laws has been guessed some time ago by Sasaki and
Yamanaka\SAYA. For this purpose we use a representation of the energy-momentum
tensor in terms of two complex fermions of spin 1, for which the dynamics
is very simple.
It turns out that the conserved densities are bilinear in the fermionic fields.
It is then a straightforward exercise to reexpress them in terms of the
energy-momentum tensor, thus recovering the explicit form of the conserved
densities.

Although the relation between singular vectors and qKdV conservation laws is
easily generalized to extended conformal algebras\DiMA\ ($N=1,2$ supersymmetry,
$W_N$, etc.) this is not so for the result described here. Nevertheless, since
the two complex spin 1 fermions provide a representation of the full $W_\infty$
algebra
\ref\LU{H. Lu, C.N. Pope, X. Shen and X.J. Wang, Phys. Lett. {\bf B267} (1991)
356.} (which is linear and thus not universal)
one could reinterpret the conserved quantities in terms of fermionic fields in
a $W_N$ context and recover a partial form of the conserved densities of the
$sl(N)$ qKdV equation (by this we mean that some, but not all, of the
coefficients of the various terms in the conserved densities are fixed). This
will be illustrated below.

At this point we are convinced that it will be very hard to go beyond these
results, that is to fully characterize in a constructive way the qKdV
conservation laws for arbitrary values of $c$.

Let us now turn to the derivation of the main result of this note.
The qKdV equation reads
\eqn\qkdv{ \dot T = [T,H] \quad,\quad H = \oint dw~(TT)(w)}
where the dot stands for a time derivative and the parentheses for normal
ordering, i.e.
\eqn\norm{ (AB)(w) = \oint {dx\over x-w} A(x)B(w). }
We will make extensive use of the following rearrangement
lemma\ref\BBSS{F.A. Bais, P. Bouwknegt, K. Schoutens and M. Surridge, Nucl.
Phys. {\bf B304}(1988) 348.}:
\eqn\rear{(A(BC)) = (B(AC)) + (([A,B])C)}
where the normal ordered commutator is easily computed from a generic OPE,
\eqn\gOPE{A(z)B(w) = \sum_r {C_r(w)\over (z-w)^r}}
to be
\eqn\NOCom{([A,B])(w) = \sum_{r>0} {(-1)^{r+1}\over r!} \partial^r C_r(w)}

At $c=-2$, $T$ can be represented by the bilinear
\eqn\T{T = (\phi\psi)}
where $\phi$ and $\psi$ are both fermions of spin 1 with OPE
\eqn\OPE{ \phi(z)\psi(w) = {-1\over (z-w)^2}\quad,\quad
\psi(z)\phi(w) = {1\over (z-w)^2}}
This is of course nothing but a ghost representation (see the last remark
at the end of this letter).
{}From the above rearrangement lemma, one easily finds that
\eqn\TT{ (TT)(z) = \frac12 (\phi''\psi + \phi\psi'')(z)}
where a prime stands for a derivative w.r.t. the complex coordinate.
The evolution of the fermionic fields can be calculated from
\eqn\dynA{\dot\phi = [\phi,H] \qquad,\qquad \dot\psi = [\psi,H]}
which yields
\eqn\dynB{\dot\phi = \phi''' \qquad,\qquad \dot\psi = \psi'''}
The dynamics of these fermionic fields is thus extremely simple.
This system of equations has an infinite number of integrals of the motion.
However, we are only interested in those that can be rewritten in terms of $T$.
A infinite set of such conserved quantities is
\eqn\IM{ H_{k+1} = \oint dz(\phi^{(k)}\psi)(z)}
where $\phi^{(k)}=\partial_z^k\phi$.
For $k$ even this can be reexpressed in terms of $T$ as follows:
\eqn\IMT{H_{2n-1} = {2^{n-1}\over n}\oint dz \Tn{n}(z)}
where the notation $\Tn{n}$ means a nesting of the normal ordering towards the
left:
\eqn\TNa{ \Tn{n} = (\dots(((TT)T)T)\dots T)\qquad (n~\hbox{factors})}
This is exactly the form of the first few conservation laws obtained in \SAYA\
for the case $c=-2$.
On the other hand, for $k$ odd the conserved integrals \IM\ cannot be expressed
in terms of $T$.

We will prove \IMT\ by showing that the exact expression for $\Tn{n}$ is
\eqn\TNb{ \Tn{n} = {n\over 2^n} \left(
\phi^{(2n-2)}\psi+\phi\psi^{(2n-2)}\right) }
for $n>1$, using a simple recursive argument.
{}From \TT, we see that \TNb\ is satisfied for $n=2$.
Let us suppose that it is true for $n$ and calculate
\eqn\TNc{ \Tn{n+1} = (\Tn{n} T) =
{n\over
2^n}\left(\left\{(\phi^{(2n-2)}\psi)+(\phi\psi^{(2n-2)})
\right\}\right)(\phi\psi) }
Let us consider the first term: $(\phi^{(2n-2)}\psi)(\phi\psi)$, to which we
apply the lemma \rear\ with $A=(\phi^{(2n-2)}\psi)$ and $B=\phi$: \eqn\eqA{
(\phi^{(2n-2)}\psi)(\phi\psi) = (\phi((\phi^{(2n-2)}\psi)\psi))
+(([(\phi^{(2n-2)}\psi),\phi])\psi) }
A further rearrangement of the first term
of the r.h.s. yields
\eqn\eqB{ (\phi^{(2n-2)}\psi)(\phi\psi) =
(\phi(\psi(\phi^{(2n-2)}\psi))) +(\phi([(\phi^{(2n-2)}\psi),\psi]))
+(([(\phi^{(2n-2)}\psi),\phi])\psi) }
Again, with some rearrangement, the first
term on the r.h.s. of the above equation can be written as
\eqn\eqC{ (\phi(\psi(\phi^{(2n-2)}\psi))) = -(\phi(\phi^{(2n-2)}(\psi\psi)))
+ (\phi(([\psi,\phi^{(2n-2)}]_+)\psi)) }
(The minus sign comes from the fermion anticommutation, and
$[A,B]_+ = AB+BA$).
Since $(\psi\psi)=0$ and $([\phi^{(n)},\psi^{(m)}]_+)=0$, this term vanishes.
The second piece in \TNc\ can be treated similarly, with the result:
\eqn\eqD{(\phi\psi^{(2n-2)})(\phi\psi) = -(([(\phi\psi^{(2n-2)}),\psi])\phi)
-(\psi([(\phi\psi^{(2n-2)}),\phi])) }
The needed normal ordered commutators are
$$\eqalign{
([(\phi^{(m)}\psi),\psi]) &= {(-1)^m\over m+2}\psi^{(m+2)} \cr
([(\phi^{(m)}\psi),\phi]) &= \frac12 \phi^{(m+2)}\cr
([(\phi\psi^{(m)}),\phi]) &= {(-1)^m\over m+2}\phi^{(m+2)} \cr
([(\phi\psi^{(m)}),\psi]) &= \frac12 \psi^{(m+2)}\cr}$$
With these results one readily obtains that
\eqn\eqE{ \Tn{n+1} = {n+1\over 2^{n+1}} (\phi^{(2n)}\psi + \phi\psi^{(2n)})}
which is exactly the form \TNb.
This recursive argument proves \TNb, whose integration establishes the
equivalence between \IM\ and \IMT.

It is straightforward to check that all the conservation laws \IM\ commute with
each other. This provides a direct proof  of the complete integrability of the
qKdV equation at $c=-2$.

Now some remarks are in order.
First, if we rearrange the conserved densities with the usual normal ordering
nested
towards the right, one finds that (except for $H_5$) the conservation laws are
in no way simpler for $c=-2$ than for any other value of $c$.
It is really the nesting towards the left which makes their expressions in
terms of $T$ simple.\foot{The symmetry breaking between left and right nesting
orders stems from the lemma \rear\ which has no simple equivalent in the other
nesting order.}
This somehow looks like an accident with no analog for the extended
cases, as far as we can see.
More explicitly, consider $(T(TT))$:
\eqn\TTT{(T(TT)) = \frac18 (\phi^{(4)}\psi + \phi\psi^{(4)}) +
\frac12 (\phi''\psi'') }
Up to a total derivative this is equal to $\Tn{3}$. However, if we introduce
another factor of $T$ on the l.h.s. of \TTT, quartic terms appear. In order
to recover the conserved density $(\phi^{(6)}\psi)$ we have to introduce
additional terms, namely $(T''(TT))$ and $(T''T'')$, reobtaining in this
way the usual form of the qKdV conserved densities of degree 8
evaluated at $c=-2$.

Secondly, it was pointed out in \LU\ that at $c=-2$, all the generators of the
$W_\infty$ algebra (which is $W_{1+\infty}$ with the spin-1 field decoupled)
can be expressed as bilinear differential polynomials in two spin-1 complex
fermions. It turns out that by considering the first $N-1$ fields as
fundamental and the remaining ones as composites,
one recovers the $W_N$ algebra. This is certainly a remarkable result given
that
$W_\infty$ is a linear algebra. It is thus natural to ask whether one could
in this way recover all the conservation laws of the $sl(N)$
qKdV equation. Unfortunately, as we will see with a few examples, this is not
the case. Nevertheless, it is interesting to notice that we do get some
information in the sense that in a candidate multi-parameter expression for a
conserved density, written in terms of the $W_N$ generators, we can fix a
certain number of these parameters simply by requiring that the conserved
quantities be of the form $(\phi^{(k)}\psi)$.

To be more explicit, let us consider the $W_3$ algebra, generated by $T$, given
in \T, and the spin-3 field $W$ whose expression in terms of the fermions
$\phi$ and $\psi$ is \LU
\eqn\W{ W = {1\over\sqrt6}(\phi'\psi-\phi\psi') }
The $sl(3)$ analog of the qKdV equation is the quantum Boussinesq
(qBsq) equation
\eqn\dynC{\dot T = [T,H] \qquad,\qquad \dot W = [W,H] }
with
\eqn\Ham{H = \oint dz~W(z) }
In terms of the fermionic fields the evolution equations (with a
trivial rescaling of the time variable) translate into
\eqn\dynD{\dot\phi = \phi'' \qquad,\qquad \dot\psi = -\psi'' }
Here again the integrals \IM\ are conserved quantities.
Now let us see whether the integrals $H_k$ for $k\not=0$ (mod 3) can be
uniquely
reexpressed in terms of $T$ and $W$ to yield the qBsq
conservation laws at $c=-2$. The first three qBsq conservation laws:
$\oint dz~T$, $\oint dz~W$ and $\oint dz~(TW)$ are readily seen to be of the
form $\oint (\phi^{(k)}\psi)$ with $k=0,1,3$.
The next one is
\eqn\IMc{\oint dz\left\{(WW) + \frac13 b^2(T(TT)) -
\frac1{60}(2-9b^2)(T'T')\right\} }
where $b^2 = 16/(22+5c)$.\foot{The corresponding expression in \KM\ has a
misprint in it.} The term $(T(TT))$ has already been calculated above (eq.
\TTT) and is equal to $\frac34 (\phi^{(4)}\psi)$, up to total derivatives.
On the other hand we have
$$\eqalign{
(WW) &= \frac16\left( (\phi'\psi)(\phi'\psi)-(\phi'\psi)(\phi\psi')-
(\phi\psi')(\phi'\psi)+(\phi\psi')(\phi\psi')\right) \cr
(T'T') &= \left( (\phi'\psi)(\phi'\psi)+(\phi'\psi)(\phi\psi')+
(\phi\psi')(\phi'\psi)+(\phi\psi')(\phi\psi')\right) \cr} $$
The second and third terms of these two expressions will produce an
non-vanishing quartic contribution $(\phi(\phi'(\psi\psi')))$ and these should
then be eliminated. This fixes the relative coefficient of the two terms
$(WW)$ and $(T'T')$, i.e.
$$ (WW) + \frac16(T'T') = \frac49 (\phi^{(4)}\psi) + \hbox{total derivative}$$
In fact, $-(2-9b^2)/60$ is indeed $1/6$ when $c=-2$.
However it is clear that the relative coefficient of $(T(TT))$ and
$(WW)+\frac16(T'T')$ is not fixed since both expressions are proportional to
$(\phi^{(4)}\psi)$ modulo total derivatives.
An analysis of the next conservation law shows that some but not all
of the coefficients are fixed.
Similarly, in the $W_4$ case, the spin-4 field (called $V$) is
proportional to \LU:
\eqn\WQUATRE{V = (\phi''\psi -3\phi'\psi' +\phi\psi'')}
The first three conservation laws \IM\ can be written as
$\oint T$, $\oint W$ and $\oint [V+\gamma(TT)]$, with $\gamma$ not fixed.

Thirdly, concerning the $W_N$ case, one could expect that there is a value of
$c=c(N)$ at which the conserved quantities become very simple when rewritten in
terms of a suitable number of ghost pairs.
{}From the point of view of Hamiltonian reduction
\ref\BER{M. Bershadsky and H. Ooguri, Comm. Math. Phys. {\bf 126} (1989) 49.}
the $W_N$ analog of $c=-2$ for $N=2$ is
\eqn\CN{c(N) = -N^4 + 2N^3 -N}
The corresponding minimal model is $(p,p')=(1,N)$. For $W_3$, this requires
a representation of $T$ and $W$ in terms of three pairs of ghosts.
However, this does not yield a simplification of the conservation laws and does
not represent a great advance when compared to the generic Feigin-Fuchs
representation in terms of two bosonic fields, valid for all values of $c$.

Let us now come back to the qKdV case.
It is well known
\ref\FMS{D. Friedan, E. Martinec and S. Shenker, Nucl. Phys. {\bf B271}
(1986) 93.}
that the energy-momentum tensor can be expressed in terms of a pair of
anticommuting ghosts $b$ and $c$ with weights $\lambda$ and $1-\lambda$, and
OPE
$$ b(z)c(w) = {1\over z-w} \qquad,\qquad c(z)b(w) = {1\over z-w} $$
as follows:
$$ T = (1-\lambda)b'c + \lambda bc' $$
The corresponding central charge is $1 - 12(\lambda-\frac12)^2$.
Previously we considered the case $\lambda=0$, with $\phi=b'$ and $c=\psi$.
Let us now see what happens when $\lambda\not=0$. Modulo a total derivative,
one finds that
\eqn\TTc{(TT) = (\frac43 \lambda(1-\lambda) + 1)(b'''c) -
2\lambda(1-\lambda)(b(b'(cc'))}
Hence for $\lambda\not=0$ (or $\lambda\not=1$, which is equivalent) $(TT)$
contains quartic terms,
The evolution equation for the $b,c$ fields takes the form (with time rescaled)
\eqn\dynE{\dot b = (\frac43 \lambda(1-\lambda) + 1)b''' +
2\lambda(1-\lambda)[2(b(b'c')) + (b(b''c))]}
and a similar equation for $\dot c$.
Unless $\lambda=0$ or 1, these are complicated coupled evolution
equations, for which \IM\ are no longer conserved quantities.
This illustrates clearly in which sense the central charge $c=-2$ is special.

\listrefs
\end